\documentclass[aps,amsmath,amssymb,12pt,showpacs,nofootinbib]{revtex4}
\usepackage[T2A]{fontenc}
\usepackage[utf8]{inputenc}
\usepackage[english,russian]{babel}
\usepackage{graphicx}
\usepackage{hyperref}

\begin{document}

\selectlanguage{english}
\title{Associated production of $J/\psi$-mesons and open charm
and double open charm production at the LHC}

\author{\firstname{A.~V.}~\surname{Berezhnoy}}
\email{Alexander.Berezhnoy@cern.ch}
\affiliation{SINP MSU, Moscow, Russia}

\author{\firstname{A.~K.}~\surname{Likhoded}}
\email{Anatolii.Likhoded@ihep.ru}
\affiliation{IHEP, Protvino, Russia}

\author{\firstname{A.~V.}~\surname{Luchinsky}}
\email{Alexey.Luchinsky@ihep.ru}
\affiliation{IHEP, Protvino, Russia}

\author{\firstname{A.~A.}~\surname{Novoselov}}
\email{Alexey.Novoselov@cern.ch}
\affiliation{IHEP, Protvino, Russia}
\affiliation{MIPT, Dolgoprudny, Russia}

\begin{abstract}
Theoretical predictions of cross sections and properties of the $J/\psi$-meson production in association with an open
charm hadron and formation of two open charm hadrons from two $c\bar{c}$ pairs in the LHC conditions 
are presented. Processes in both single and in double parton scattering mechanisms are included 
into consideration. Special attention is paid to the kinematic limits of the LHCb detector for which comparison
with the newest experimental data is carried out.
\end{abstract}

\pacs{13.85.Fb, 14.40.Rt}

\maketitle

\section{Introduction}

In recent work of the LHCb collaboration~\cite{Aaij:2011yc} data on the double $J/\psi$-meson
production at $7~\mathrm{TeV}$ energy is presented.
At first blush value of the double $J/\psi$ production cross section reported ($5.1\pm1.1~\mathrm{nb}$) is
accordant within uncertainty limits with the predictions obtained in
the leading order (LO) QCD calculations~\cite{Berezhnoy:2011xy, Baranov:2011zz}. 
These calculations lead to the total cross section value of $10 \div 27~\mathrm{nb}$
and to $3 \div 5~\mathrm{nb}$ in the kinematic limits of the LHCb detector.
It is well known that such calculations include big uncertainties connected with the hard scale 
selection, next to the LO (NLO) contributions and allowing for the relative motion of $c$-quarks in
the $J/\psi$-meson. 
It is known that the last of this factors increases cross section of double quarkonia production
in $e^+e^-$-annihilation in several times~\cite{Berezhnoy:2007sp,Braguta:2005kr,Braguta:2007ge,Braguta:2008hs}.

Apart from uncertainties in the partonic cross section of double $J/\psi$ production a new problem arises in 
the LHC conditions. The hadronic cross section appears to be three orders of magnitude higher than the cross section
of the partonic subprocess.
This phenomenon dues to the high luminosity of low-$x$ gluons with fraction of proton momenta 
of about $10^{-4}\div10^{-3}$, which contribute most to the processes in question.
Such an enhancement gave rise to discussion of double parton interactions in a single $pp$-collision 
(DPS) \cite{Paver:1982yp, Flensburg:2011kj, Diehl:2011yj, Bartalini:2011jp} with independent production
of particles considered in each interaction. 
In works~\cite{Baranov:2011ch, Novoselov:2011ff, Kom:2011bd} mechanism of double $J/\psi$ production
in DPS approach was considered and it was shown that DPS can give significant contribution to the channel in 
question in the LHCb detector conditions.

Although SPS\footnote{We will address formation of considered final states in a single parton interaction as SPS.}
and DPS models predict somewhat different kinematic distributions for the $J/\psi$ pairs
produced, the question if enhancement of statistics gained allows to distinguish this mechanisms 
remains open.
On the other hand at least in the LO there is a qualitative
difference between predictions for the $J/\psi+\chi_c$ production obtained in the SPS and DPS models.

Moreover, additional DPS contribution should obviously express itself in other channels of the four $c$-quark domain:
in the associated production of $J/\psi+D$\footnote{In the following we will refer to the $J/\psi+D$ production for the production of 
$J/\psi$ and a $c\bar{c}$ pair, from which at least one $c$-quark hadronize into an observed open charm hadron.}
and in the four $D$-meson production\footnote{In the following we will refer to the four $D$-meson production for the production of 
a $c\bar{c}c\bar{c}$ configuration, from which at least two $c$-quarks hadronize into observed open charm hadrons.}.
In the beginning of 2012 first LHCb results for the channels listed were presented~\cite{LHCb-PAPER-2012-003}.
It is interesting to understand the interplay of the SPS and DPS mechanisms in these channels.
Currently there are estimations of cross sections of the SPS processes contributing these final states in the LO perturbative QCD
formalism~\cite{Barger:1991vn, Berezhnoy:1998aa, Baranov:2006dh, Baranov:2006rz, Artoisenet:2007xi, He:2009zzb}. 
In the current work we review results obtained in the LO perturbative QCD for the SPS contribution and estimate DPS contributions
for the channels mentioned.

\section{Four heavy quark production in the single gluon-gluon interaction}

One of the first research, in which the possibility to observe four heavy quark production at colliders 
was discussed is~\cite{Barger:1991vn}. 
In this work cross sections of subprocesses  $gg \to Q_1\bar{Q}_1 Q_2 \bar{Q}_2$ and $q \bar{q}\to Q_1\bar{Q}_1 Q_2 \bar{Q}_2$ 
was estimated within LO perturbative QCD approach for the kinematical conditions of the LHC and SSC.  

Slightly later the analogous processes in which quark and antiquark are bind in a doubly-heavy meson 
were investigated.     
Production of the $S$-wave $B_c$-meson in the $gg \to B_c + b + \bar{c}$ and $q \bar{q}\to B_c + b + \bar{c}$ processes
was estimated in the works~\cite{Berezhnoy:1994ba, Berezhnoy:1995au, Berezhnoy:1997fp, Kolodziej:1995nv, Chang:1994aw, 
Baranov:1997sg, Baranov:1997wv, Chang:2005wd}.
Calculation of the $P$-wave $B_c$-meson production cross section was done in the 
studies~\cite{Berezhnoy:1996ks, Berezhnoy:1997uz, Chang:2004bh}.
These researches continued in the works~\cite{Berezhnoy:2004gc, Chang:2006xka, Berezhnoy:2010wa}.
Associated $J/\psi$ and $D$-meson production, as well as $\Upsilon$ and $B$-meson production, was also
estimated within the same technique in~\cite{Berezhnoy:1998aa, Baranov:2006dh, Baranov:2006rz, Artoisenet:2007xi, He:2009zzb}.

It is worth to note that doubly heavy baryon production implies production of two heavy quarks. 
Therefore, assuming that the doubly heavy baryon is created in the heavy diquark hadronization, 
one can study the doubly heavy baryon production by analogy with the $B_c$-meson 
production~\cite{Baranov:1997xn, Berezhnoy:1995fy,  Berezhnoy:1998aa, Chang:2006eu, Zhang:2011hi}. 
 
Calculations show that gluon-gluon interactions provide the main contribution into the four heavy quark production 
in the LHC experiments.
Quark-antiquark annihilation amounts to about 10\%.
That is why production in the gluon-gluon interactions is mainly discussed in this paper.   

Usually calculations are made under an assumption that initial gluons are real and their transverse momenta 
are negligible (the collinear approach).
To simulate real distribution over the transverse momenta of initial gluons in our studies we use the 
Pythia 6.4 MC generator~\cite{Sjostrand:2006za}.
In this connection it is worth noting researches~\cite{Baranov:2006dh,Baranov:2006rz}
where transverse momenta and virtualities of gluons are taken into account in the framework of the $k_T$-factorization approach. 

\section{Pair production of $J/\psi$-mesons in the LHCb detector}

Production of two charmonia in SPS can be described within perturbative QCD by the fourth order in $\alpha_S$
Feynman 
diagrams. For the $J/\psi$-mesons pair formation invariant masses and quantum numbers ($1^{--}$) 
of two $c\bar{c}$ pairs are fixed.

Cross section of the hard subprocess of two $c\bar{c}$ pair formation in the color-singlet ($1_\mathrm{C}$) state with 
$m_{c\bar{c}}\approx m_{J/\psi}$ is proportional 
to\footnote{For the precise expression see~\cite{Humpert:1983yj,Ko:2010xy,Berezhnoy:2011xy}.}
\begin{equation}
\hat{\sigma}(gg \to J/\psi J/\psi)\sim\frac{\alpha_S^4 |\psi(0)|^4}{m_{J/\psi}^8},
\label{eqn:sigJJ}
\end{equation}
where $|\psi(0)|$ is the value of the $c\bar{c}$ wave function in the $J/\psi$-meson at the origin.
Emergence of this factor dues to the approximation in which momenta of $c$ and $\bar{c}$ quarks are 
parallel and their relative momentum is neglected in the matrix element of the subprocess in question
($\delta$-approximation).
At large invariant masses of the $J/\psi$-meson pair cross section (\ref{eqn:sigJJ}) decreases
with the rise of the full energy squared $\hat{s}$ as
\begin{equation}
\hat{\sigma}(gg \to J/\psi J/\psi)\sim\frac{\alpha_S^4 |\psi(0)|^4}{\hat{s}^4}.
\label{eqn:sigJJs}
\end{equation}

Numerical result of $4.1\pm1.2~\mathrm{nb}$~\cite{Berezhnoy:2011xy} derived in the assumptions listed 
was obtained using the hard subprocess scale equal to the transverse mass of one of the $J/\psi$-mesons
produced and using the CTEQ5L proton \textit{pdf}s~\cite{Lai:1999wy}. 
As mentioned above this value agrees within uncertainty limit with the experimental value of $5.1\pm1.1~\mathrm{nb}$
measured in~\cite{Aaij:2011yc}.
Variation of the hard scale from the one half to two transverse masses of the $J/\psi$-meson produced
changes the cross section value from $5.1~\mathrm{nb}$ to $3~\mathrm{nb}$.
If CTEQ6LL \textit{pdf}s~\cite{Pumplin:2002vw} are used, cross section has maximum at the scale of about one
transverse mass of $J/\psi$ and amounts to $3.2~\mathrm{nb}$. 
Cross sections are less at both half and double scales and are $2.8~\mathrm{nb}$ and $2.6~\mathrm{nb}$ respectively.
Presence of extremum dues to the fact that with rise of the scale strong coupling constant decreases while
gluon density grows.
As well as in the manuscript~\cite{Berezhnoy:2011xy} we include contribution from the production and decay
of the $\psi(2S)$ state into the $J/\psi$-mesons yield.

\begin{figure}
\includegraphics[width=12cm]{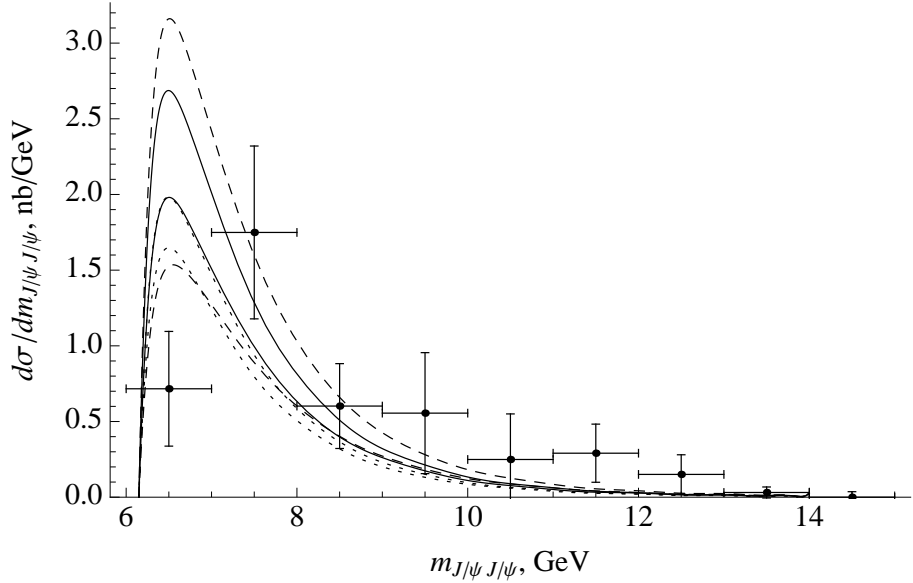} 
\caption{Distribution over the invariant mass of the $J/\psi$-meson pairs compared with the LHCb measurement. 
Solid curves were obtained with $m_T^{J/\psi}$ as the hard scale, dashed --- $2 \cdot m_T^{J/\psi}$ and dotted ---
$0.5 \cdot m_T^{J/\psi}$. For every scale choice upper curve corresponds to the CTEQ5L, lower ---  to the CTEQ6LL \textit{pdf}
used.}.
\label{fig:exp2psicmp}
\end{figure}

Fig.~\ref{fig:exp2psicmp} 
shows distributions over the invariant mass of the $J/\psi$ pairs calculated within the assumptions mentioned
for the different hard scale choices and \textit{pdf} sets in comparison with the experimental data reported by LHCb 
in ~\cite{Aaij:2011yc}.
One can see that the shape of the distributions predicted is nearly the same.
What concerns experimental distribution, it looks tilted to the bigger invariant mass values. 
We would like to notice that this fact can be accounted for by the relative $c$-quarks motion in the $J/\psi$-meson.
With this aim we calculated cross section of the process in question averaged by some
``duality'' region of the $c\bar{c}$ invariant mass:
\begin{multline}
\hat \sigma^{\rm{dual}} (gg \to J/\psi (\psi') J/\psi(\psi')) \approx \\ \approx \int\!\!\! \int_{2m_c}^{2m_{D}+\Delta} 
\frac{d^2\sigma\left( gg \to (c\bar c)^{S=1}_{\rm{1_C}}+(c\bar
c)^{S=1}_{\rm{1_C}}\right)}{dm_{{c\bar c}_1}dm_{{c\bar c}_2}}
dm_{{c\bar c}_1}dm_{{c\bar c}_2},
\label{eqn:2jpsi_dual}
\end{multline}
where $m_D$ is the $D$-meson mass. The $c$-quark mass was taken equal to
$$m_c=1.25 \mbox{ GeV}.$$
The $\Delta$ parameter can be selected in such a way that the value of the pair production cross section obtained 
coincides with the total production cross section of the $J/\psi J/\psi$, $J/\psi \psi'$ and $\psi' \psi'$ final states
calculated in the $\delta$-approximation.
If one takes $\sqrt{\hat s}/2$ for the hard scale of the subprocesses considered the correspondence is reached at  
$\Delta=0.3 \mbox{ GeV}$:
$$\hat \sigma^{\rm{dual}} (gg \to J/\psi (\psi') J/\psi(\psi'), \Delta=0.3 {\mbox{ GeV}} ) \approx 4.4 \mbox{ nb.}$$

At $\Delta=0.5 \mbox{ GeV}$ the cross section estimated in the duality approach is close to the value reported by the LHCb 
experiment: 
$$\sigma^{\rm{dual}} (gg \to J/\psi (\psi') J/\psi(\psi'), \Delta=0.5 {\mbox{ GeV}}) \approx 5.8 \mbox{ nb.}$$

\begin{figure}
\includegraphics[width=12cm]{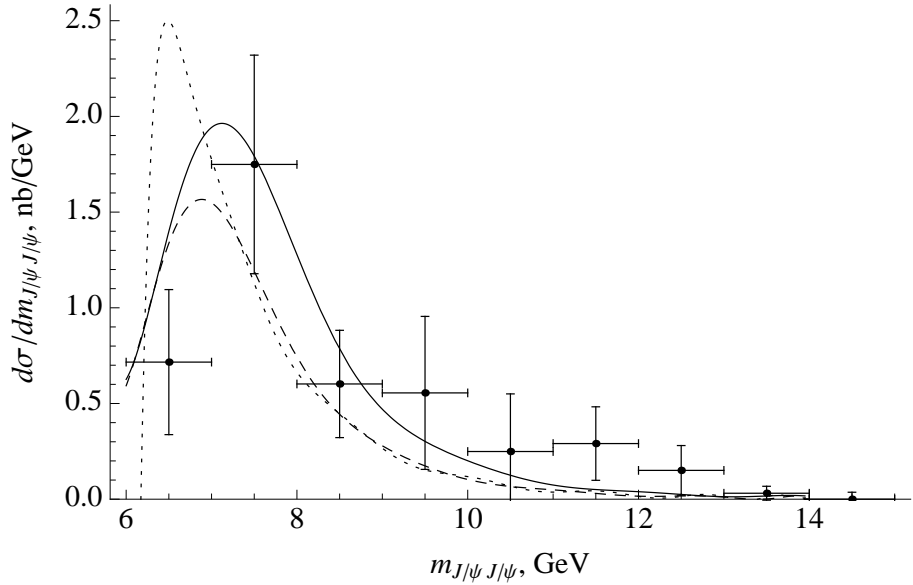} 
\caption{Distribution over the invariant mass of the $J/\psi$-meson pairs in the ``duality'' approach
compared with the LHCb measurement. Solid curve was obtained with $\Delta=0.5$~GeV, dashed --- 
with $\Delta=0.3$~GeV and dotted --- in the $\delta$-approximation.}
\label{fig:duality_psi}
\end{figure}

Increase of $\Delta$ leads to the growth of the total cross section on the one hand, and improves 
agreement in the $m_{J/\psi J/\psi}$ distribution on the other (Fig.~\ref{fig:duality_psi}).

In the LHC environment huge density of low-$x$ gluons leads to the increase of the multiple gluon-gluon interactions probability
within one proton-proton collision.
In the DPS approach, which implies production of particles concerned in two independent subprocesses,
the cross section is written down as following:
\begin{equation} 
\label{eqn:doubleAB}
\sigma^{ A B }_{\rm DPS} = \frac{m}{2} \frac{\sigma^{ A}_{\rm SPS} 
\sigma^{ B}_{\rm SPS}} {\sigma_{\rm eff}}.
\end{equation} 
where the $\sigma_{\rm eff}=14.5~\mathrm{mb}$ parameter was measured in the four jets and three jets plus photon
modes by the CDF and D0 detectors \cite{Abe:1997xk,Abazov:2009gc}. 
The $m$ parameter equals $1$ for identical subprocesses and $2$ for different.
For the $J/\psi$ pairs production in the LHCb conditions expression (\ref{eqn:doubleAB}) leads to
the following cross section value:
\begin{eqnarray} 
\sigma_{\rm DPS}^{pp\to J/\psi J/\psi+X} &=& 4~\mathrm{nb}.
\end{eqnarray} 
Known inclusive production cross section of the $J/\psi$ meson in the LHCb kinematic limits,
$\sigma_{J/\psi} = 10.5~\mathrm{\mu b}$, was used. 
In the work~\cite{Kom:2011bd} authors note that the DPS contribution can be located at bigger
$J/\psi$ pair invariant masses than the SPS one.

\begin{figure}
\includegraphics[width=12cm]{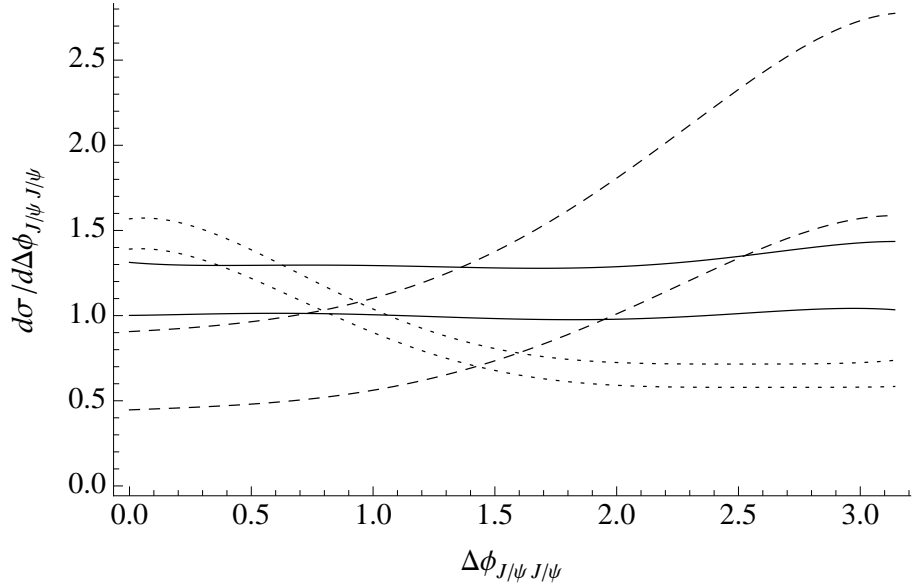} 
\caption{Distribution over the difference of the azimuthal angles of the $J/\psi$-mesons produced. Curve designations
coincide with Fig. \ref{fig:exp2psicmp}.}
\label{fig:DphiJJ}
\end{figure}

\begin{figure}
\includegraphics[width=12cm]{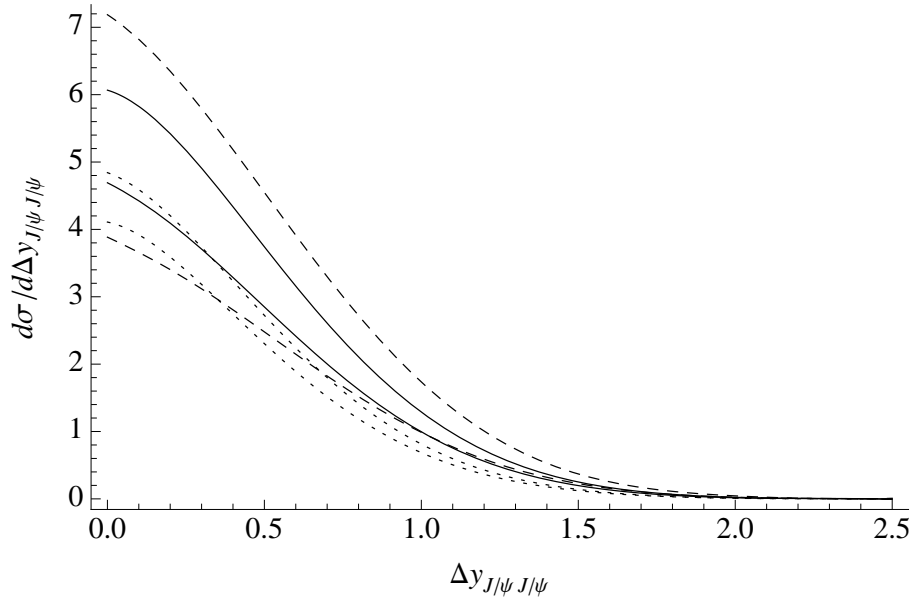} 
\caption{Distribution over the difference of rapidities of the $J/\psi$-mesons produced. Curve designations
coincide with Fig. \ref{fig:exp2psicmp}.}
\label{fig:DyJJ}
\end{figure}

One of the proposed methods to distinguish the DPS signal from the SPS one is to study correlations
between azimuthal angles of two mesons produced or between their rapidities~\cite{Kom:2011bd,Novoselov:2011ff}.
However analysis involving modelling in the Pythia generator shows that correlations presenting in collinear 
approach completely go out when including transverse momenta of the initial gluons into consideration (Fig.~\ref{fig:DphiJJ}).
To be more precise, depending on the scale choice collinear or anticollinear directions of the $J/\psi$ momenta can dominate.
Moreover, at the standard scale of one $J/\psi$ transverse mass the relative angle correlation is absent at all.
It is the model implemented in the Pythia generator which is completely responsible for the distribution over
the transverse momenta of the initial gluons, so model-independent prediction on the $\Delta \phi$
distribution can not be derived.
Investigation of the rapidity correlation appears to be more fruitful.
In spite of narrowness of LHCb rapidity window ($2.0<y<4.5$) it appears to be sufficient
to test QCD predictions which state that the difference in rapidity between the $J/\psi$-mesons 
produced does not exceed $2$ units of rapidity (Fig.~\ref{fig:DyJJ}).
At the current stage DPS predicts no correlations between products of two partonic interactions at all.

Apart from the correlation studies investigation of the $P$-wave state contributions to the total
$J/\psi$ production can be fruitful.
Indeed, section rules emerging in the CS LO pQCD consideration \cite{Kartvelishvili:1984ur} imply significant limitation
on the final states which can appear in the gluon fusion process.
According to the $C$-parity conservation occurrence of the $C$-odd $J/\psi\chi_C$ final state 
should be suppressed in the SPS. 
As in DPS formation of charmonia occurs independently, it does not have any suppression in this channel.
That is why DPS should dominate in the $J/\psi \chi_C$ state production (possibly followed by 
the $\chi_C \to J/\psi \gamma$ decay).
It was pointed out in the work~\cite{Novoselov:2011ff} that similar situation takes place in the $J/\psi\Upsilon$
mode in which SPS and DPS lead to different hierarchy of the 
$pp \to J/\psi J/\psi$, $pp \to  J/\psi \Upsilon$ and $pp \to \Upsilon \Upsilon$ cross sections.

As it was noted, we have taken into account contribution of the $pp \to J/\psi \psi(2S)$ and $pp \to \psi(2S) \psi(2S)$ processes
followed by the $\psi(2S) \to J/\psi X$ decays to the total $J/\psi$ pairs yield. $\psi(2S)$-mesons originating from these processes
can also be detected by a leptonic decay, just like $J/\psi$. It is interesting to compare ratios of $J/\psi J/\psi$ and $J/\psi \psi(2S)$
yields predicted by SPS and DPS models.

Both in SPS and in DPS approaches ratio of different meson pair yields can be estimated using values of the $c\bar{c}$
wave functions in the charmonia at the origin:
\begin{eqnarray}
&\sigma(pp \to J/\psi J/\psi) : \sigma(pp \to J/\psi \psi(2S)) : \sigma(pp \to \psi(2S) \psi(2S))& \approx
\nonumber \\
\approx &\psi_{J/\psi}(0)^4 : 2 \cdot \psi_{J/\psi}(0)^2 \psi_{\psi(2S)}(0)^2 : \psi_{\psi(2S)}(0)^4& \approx
\nonumber \\
\approx &1 : 1 : 0.3.&
\end{eqnarray}
where $\psi_{J/\psi}(0) = 0.21~\mathrm{GeV}^{3/2}$, $\psi_{\psi(2S)}(0) = 0.16~\mathrm{GeV}^{3/2}$.
A more accurate estimate which allows for different meson masses leads to the relation
\begin{eqnarray}
&\sigma(pp \to J/\psi J/\psi) : \sigma(pp \to J/\psi \psi(2S)) : \sigma(pp \to \psi(2S) \psi(2S))& \approx
\nonumber \\
\approx &1.7 : 1 : 0.15.&
\end{eqnarray}
Accounting contributions from the $\psi(2S)$ decays in the channels discussed one gets finally
\begin{eqnarray}
&\sigma(pp \to J/\psi J/\psi) : \sigma(pp \to J/\psi \psi(2S)) : \sigma(pp \to \psi(2S) \psi(2S))& \approx
\nonumber \\
\approx &2.2 : 1 : 0.13.&
\label{eqn:SPSwfd}
\end{eqnarray}
What concerns DPS, using inclusive $J/\psi$ and $\psi(2S)$ production cross sections equal to 
$10.5~\mathrm{\mu b}$~\cite{LHCb-CONF-2010-013} and 
$1.88~\mathrm{\mu b}$~\cite{LHCb-CONF-2011-026} respectively, one gets
\begin{eqnarray}
&\sigma(pp \to J/\psi J/\psi) : \sigma(pp \to J/\psi \psi(2S)) : \sigma(pp \to \psi(2S) \psi(2S))& =
\nonumber \\
= &\sigma_{J/\psi}^2 : 2 \cdot \sigma_{J/\psi} \sigma_{\psi(2S)} : \sigma_{\psi(2S)}^2& =
\nonumber \\
= &2.8 : 1 : 0.9.&
\label{eqn:DPSwchiC}
\end{eqnarray}
It can be seen that DPS predicts slightly larger suppression of the $J/\psi\psi(2S)$ production compared to SPS.
The main reason of it is that inclusive $J/\psi$ production cross section already includes contribution from the 
$\chi_C$ decays, which can amount up to $20 \div 30\%$ \cite{Abe:1997yz, Abt:2008ed}. If one excludes expected
contribution of $\chi_C$ decays by taking $J/\psi$ production cross section equal 
$0.8 \times 10.5~\mathrm{\mu b}= 8.4~\mathrm{\mu b}$, then DPS prediction amounts to
\begin{eqnarray}
&\sigma(pp \to J/\psi J/\psi) : \sigma(pp \to J/\psi \psi(2S)) : \sigma(pp \to \psi(2S) \psi(2S))& =
\nonumber \\
= &\sigma_{J/\psi}^2 : 2 \cdot \sigma_{J/\psi} \sigma_{\psi(2S)}^2 : \sigma_{\psi(2S)}^2& =
\nonumber \\
= &2.2 : 1 : 0.11.& 
\label{eqn:DPSwochiC}
\end{eqnarray}
Up to the uncertainties this relation coincides with the SPS prediction (\ref{eqn:SPSwfd}). Uncertainties in the cross section ratios 
predicted by DPS can be estimated by the largest relative uncertainty in the measurement of the cross sections involved. 
This uncertainty is maximum for the $\psi(2S)$ measurement and reaches $20\%$~\cite{LHCb-CONF-2011-026}. 
Unfortunately difference between 
relation (\ref{eqn:DPSwchiC}), which suspects feed-down from the $\chi_C$ production, and relations (\ref{eqn:SPSwfd}), (\ref{eqn:DPSwochiC}),
which do not, is of the same order. Nonetheless it would be interesting to measure ratio of the $J/\psi J/\psi$ and $J/\psi\psi(2S)$ 
yields experimentally.

\section{Associated production of $J/\psi$ and $D$ meson in the LHCb detector}

To compare predictions for the $gg \to J/\psi c \bar  c$ and 
$gg \to c \bar c  c \bar c$ processes with experiment some model of the $c$-quark 
transition into a specific hadron should be used.
The most common hardonization model is based on the assumption that  
charm hadron moves approximately in the same direction that the  initial $c$ quark does
and obtains some fraction $z$ of the quark momentum with the probability $D_{c \to H}(z)$ (so called fragmentation function).
At the scale of about $c$-quark mass the mean $z$ value is about 0.7.    
Two following parametrizations are used in our calculations: 
the standard parametrization of Pythia 6.4 and a pQCD motivated parametrization of BCFY~\cite{Braaten:1994bz} 
with the parameter values obtained in~\cite{Cacciari:2003zu}.

It is worth mentioning here that as it was shown 
in~\cite{Berezhnoy:2000ji,Berezhnoy:2010zz,Berezhnoy:2005dt,Braaten:2001bf,Braaten:2001uu},
there are models in which hadronization is not described by simple fragmentation.
For example, it is reasonable to suppose that $c$-quark can pull a light quark from the sea without loosing any momentum. 
In this case it can be formally assumed that $D_{c \to H}(z)=\delta(z)$.
Moreover it can be supposed that in some cases the final hadron momentum is even larger 
(by a quantity of about $\frac{m_q}{m_c}p_c$) than the initial $c$-quark momentum. 

All mentioned possibilities have been  considered in the present estimations of the cross section values. 
Nevertheless it should be stressed that these estimations are too rough to give preference to some particular hadronization model.

Recently cross section value of the associated production of $J/\psi$ together with a $D$-meson has been measured by
the LHCb collaboration for the following kinematical region~\cite{LHCb-PAPER-2012-003}: 
\begin{itemize}
\item $J/\psi$ meson is produced in the rapidity region $2.0<y_{J/\psi}<4.0$;
\item one charmed hadron is produced in rapidity region $2.0<y_{D}<4.0$ and has transverse momenta 
$3 \mbox{ GeV} < p_T^D < 12 \mbox{ GeV}$.
\end{itemize}

Calculations within the LO of pQCD lead to the cross section value $20\div 60 \mbox{ nb}$ depending on 
the scale choice and the $c$-quark hadronization model~\cite{Berezhnoy:1998aa, Baranov:2006dh, Artoisenet:2007xi, He:2009zzb}
(the scale value was varied from $\sqrt{\hat s}/4$ to $\sqrt{\hat s}$).
Nevertheless, as it was shown is paper~\cite{Baranov:1997sg}, interaction of the sea $c$-quark from one proton 
with gluon from the other can essentially contribute to the  $J/\psi$-meson and $c$-quark associated production, 
i.e. the subprocess $cg \to J/\psi c$\footnote{Form now on summation with charge conjugate mode is implied.} 
should also be taken into account, as well as the main subprocess $gg \to J/\psi c \bar c$. 
It is natural for such an approach, that problems connected with double counting and non-zero $c$-quark mass 
essentially impede an accurate estimation of the calculation uncertainties. 
It can be assumed that this method is already valid at the transverse momenta of the charmed hadron 
$p_T^D>3\mbox{ GeV}\approx 2m_c$ and that interference contributions are small.
Also one can try to avoid double counting by subtracting the part due to the direct gluon splitting
from the total $c$-quark structure function: 
\begin{equation}
\tilde f_c(x,Q^2)=f_c(x,Q^2)-\frac{\alpha_s(Q^2)}{4\pi}\int_x^1 \frac{dz}{z}
\left[\Bigl(\frac{x}{z}\Bigr)^2+\Bigl(1-\frac{x}{z}\Bigr)^2 + \frac{2m_c^2 x (z-x)}{z^2Q^2}\right] f_g(z,Q^2),
\label{eqn:c_pdf}
\end{equation}  
where splitting function is taken from~\cite{Martin:1996eva}.

The cross section value of the subprocesses $cg \to J/\psi c$ was found to be about $10\div 40~\mathrm{nb}$.
Therefore the contribution of such corrections to the $J/\psi+D$ associated production is of the same order as 
the contribution of the main subprocess $gg\to J/\psi c \bar{c}$. 

Thus the calculations within pQCD lead to the cross section value of about $30\div 100$~nb for the $J/\psi+D$ production
in the LHCb fiducial region. It should be noticed that in contrast to the charmonia pairs production in the 
associated charm production there are no $C$-parity selection rules. So one should expect not only feed-down from
the $J/\psi+\psi(2S)$ production but also from the $J/\psi+\chi_C$ one. This contributions can increase observed 
$J/\psi + D$ cross section by up to 50\%.

\begin{figure}
\includegraphics[width=12cm]{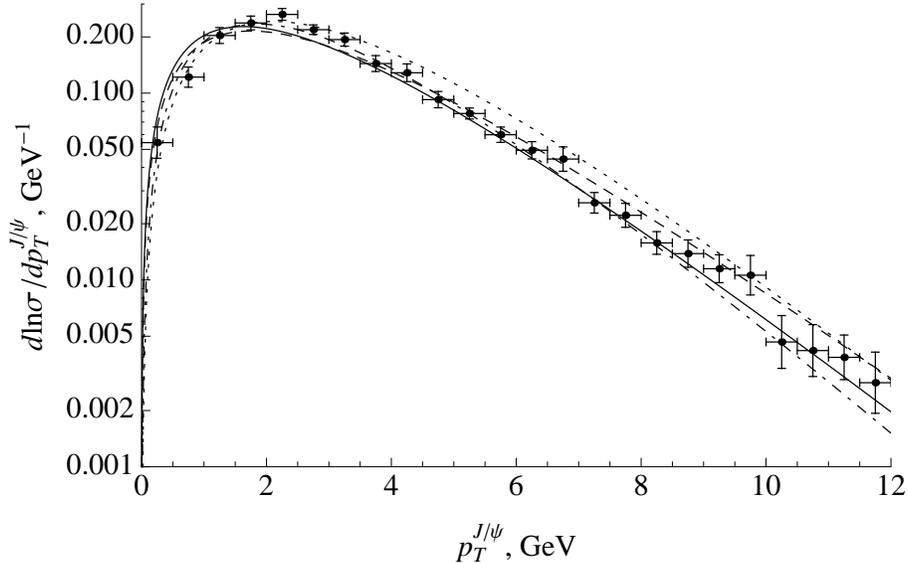} 
\caption{Distribution over the transverse momentum of the $J/\psi$-meson in the $J/\psi + D$ production 
compared with the LHCb measurement (points for $J/\psi$ produced together with $D^0$ or $D^+$-meson are shown). 
Solid curves were obtained with the hard scale value of $1\cdot m_T^{J/\psi}$, dashed --- $2 \cdot m_T^{J/\psi}$ and dotted ---
$0.5 \cdot m_T^{J/\psi}$. Dot-dashed curve corresponds to the collinear gluon approximation.}
\label{fig:PsiPtExpQCD}
\end{figure}

In the Fig.~\ref{fig:PsiPtExpQCD} 
cross section distribution over the $J/\psi$ meson transverse momentum in 
the $gg \to J/\psi+c\bar{c}$ subprocess is shown in comparison with the LHCb experimental data. 
The $d \ln\sigma/d p_T$ distributions are plotted, i.e. spectra are normalized to unity. 
Both $J/\psi$ and associated charmed hadron produced in the events plotted are limited to the LHCb fiducial region.
It can be seen that at least in the high $p_T^{J/\psi}$ region the predicted slope is in a good agreement with the
experimental data. It should be noticed that $p_T^{J/\psi}$ distribution in the inclusive $J/\psi$ production 
measured by LHCb exhibits significantly more rapid decrease with the $p_T^{J/\psi}$ growth.

\begin{figure}
\includegraphics[width=12cm]{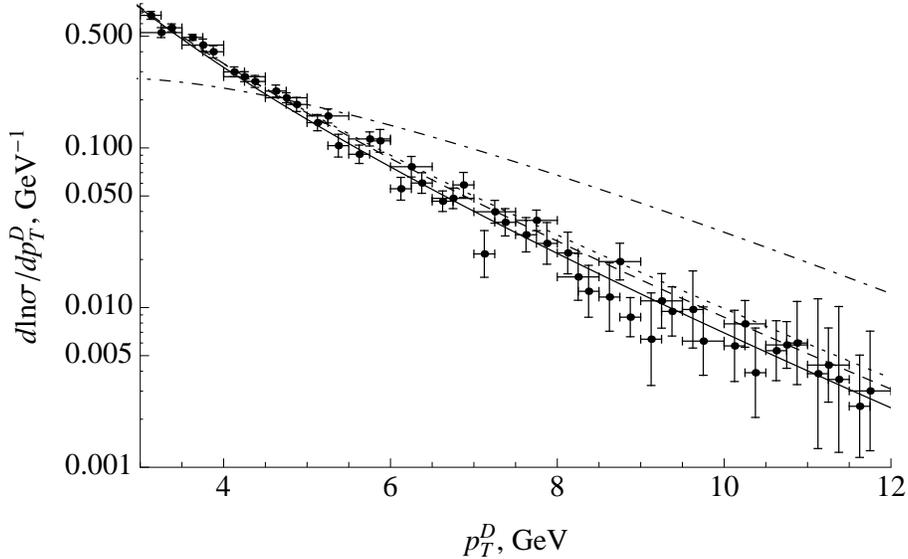} 
\caption{Distribution over the transverse momentum of the $D$-meson in the $J/\psi + D$ production 
compared with the LHCb measurement (points for $D^0$ and $D^+$-mesons are shown). 
Curve designations coincide with Fig. \ref{fig:PsiPtExpQCD}.}
\label{fig:DPtExpQCD}
\end{figure}

Cross section distribution over the $D$-meson transverse momentum for the same
$gg \to J/\psi+c\bar{c}$ subprocess is given in the Fig.~\ref{fig:DPtExpQCD}
and demonstrates good agreement with the LHCb measurement. As in the previous figure, both spectra are normalized to unity. 
In contrast to the $J/\psi$ signal, both predicted and measured spectra are similar to those in 
the inclusive $D$-meson production at LHCb~\cite{LHCb-CONF-2010-013}.

\begin{figure}
\includegraphics[width=12cm]{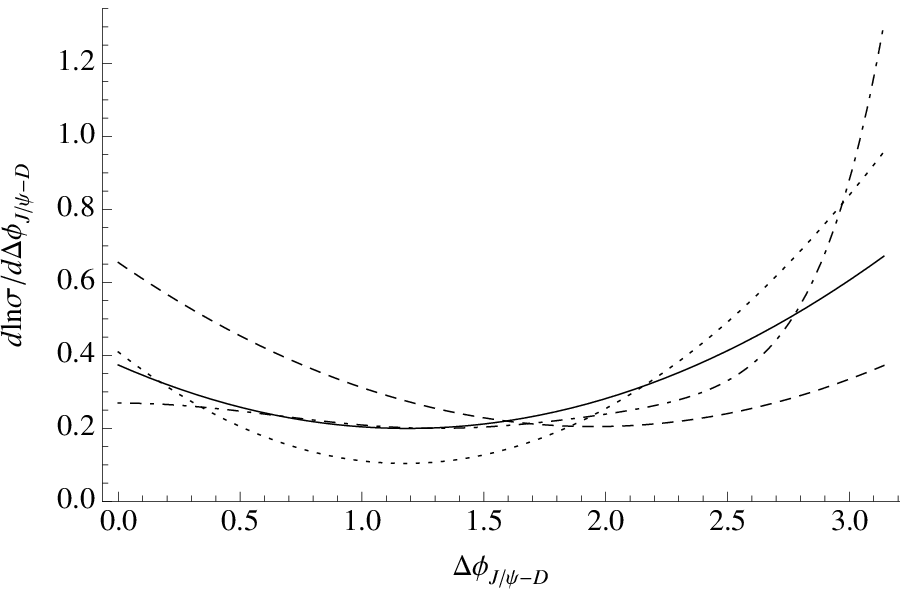} 
\caption{Distribution over the difference of the azimuthal angles of the $J/\psi$ and $D$-meson in the $J/\psi + D$ production.
Curve designations coincide with Fig. \ref{fig:PsiPtExpQCD}.}
\label{fig:DphiJD}
\end{figure}

As in double $J/\psi$ production, essential angle and rapidity correlations in the
$gg \to J/\psi c \bar{c}$ process are predicted by pQCD.
Within collinear approach $J/\psi$ and $D$ mesons move in the opposite directions in most cases.
However no concrete prediction can be made 
when taking into account transverse gluon motion in the framework of the
Pythia generator as the distribution is highly sensitive on the scale selection (see Fig.~\ref{fig:DphiJD}).

\begin{figure}
\includegraphics[width=12cm]{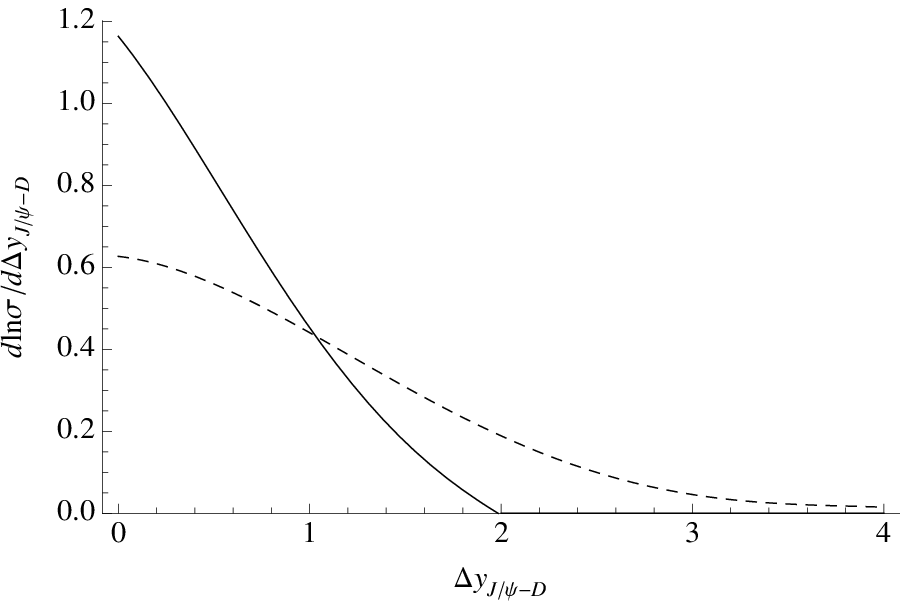} 
\caption{Distribution over the difference of rapidities of the $J/\psi$ and $D$-meson in the $J/\psi + D$ production.
Solid curve corresponds to the LHCb kinematic limits imposed, dashed --- to the absence of kinematic limits.}
\label{fig:DyJD}
\end{figure}

What concerns distribution over the rapidity difference between $J/\psi$ and $D$-meson produced, from Fig.~\ref{fig:DyJD} 
one can see that contrary to the two $J/\psi$-meson production, LHCb rapidity window appears to be too narrow
to observe rapidity correlations predicted in the $gg \to J/\psi c \bar{c}$ subprocess. We omit discussion of correlations
between $D$ and $\bar{D}$ mesons in the $J/\psi + D$ associated production as LHCb analysis focuses on events in which
one co-produced $D$ meson is observed.

The cross section value obtained by LHCb collaboration,
\begin{equation}
\sigma^\mathrm{exp}(pp \to J/\psi + D^0(D^+,D^+_s,\Lambda^+_c)+X)\approx 300 \mbox{ nb},
\label{eqn:PsiDexp}
\end{equation}
is several times larger than the SPS prediction of $30\div 100~\mathrm{nb}$.

Let us now address to the simultaneous production of $J/\psi$ and open charm in two gluon-gluon interactions.
Within the DPS approach cross section of the associated $J/\psi$ and $D$-meson production can be expressed as follows:
\begin{equation}
\sigma^{\rm DPS}_{J/\psi D} = \frac{\sigma_{J/\psi} \sigma_D}{\sigma_{\rm eff.}},
\end{equation}
where $\sigma_{J/\psi}$ and $\sigma_D$ are cross sections of the inclusive $J/\psi$ and $D$-meson production in the 
LHCb acceptance correspondingly. 
Recalculated for the fiducial region discussed ($2<y<4$, $p_T^D>3~\mathrm{GeV}$), these values are 
$9~\mathrm{\mu b}$ and $380~\mathrm{\mu b}$ respectively~\cite{Aaij:2011jh,LHCb-CONF-2010-013}. 
As always summation with the charge conjugate state is assumed.
Unpublished cross section of the $\Lambda_C$ inclusive production is not included in consideration.
Thus the associated $J/\psi$ and $D$ meson production cross section for the LHCb kinematical region 
within the DPS model can be estimated as 
\begin{equation}
\sigma^{\rm DPS}_{J/\psi D} = 240~\mathrm{nb}.
\end{equation}
As earlier, numerical value of $\sigma_{\rm eff.}=14.5~\mathrm{mb}$~\cite{Abe:1997xk,Abazov:2009gc} was used.
One can see that DPS prediction is several times larger than the SPS one and within uncertainty limits agrees 
with the experimental value~(\ref{eqn:PsiDexp}).

\section{Four $D$-meson production in the LHCb detector}

In the same LHCb studies~\cite{LHCb-PAPER-2012-003} production of four $c$-quarks is investigated. 
Events in which two open charm hadrons both containing $c$-quark (or both containing $\bar{c}$-quark)
are produced in the fiducial region $2.0<y<4.0$, $3 \mbox{ GeV} < p_T < 12 \mbox{ GeV}$ were selected.

The calculation within LO of QCD in SPS approach gives
for this kinematical region cross section value of
$$\sigma^\mathrm{pQCD} (gg\to c \bar c c \bar c)\sim 50 \div 500~\mathrm{nb}$$
depending on the scale selection and the $c$-quark hadronization model used. 

There is an indication that interaction with sea $c$-quarks contribute essentially into this process, 
as well as into the associated production of $J/\psi$ and $c$.
According to our preliminary estimation, cross section value for the process  $c g \to c c\bar c$ 
(plus charge conjugate) is about
 $$\sigma^\mathrm{pQCD} (c g\to  c c \bar c) \sim 200 \div 500~\mathrm{nb}$$
depending on the scale selection and the $c$-quark hadronization model used. 
The $c$-quark structure function has been taken in the form~(\ref{eqn:c_pdf}). 

The interactions between two sea $c$-quarks can also be considered.
Our estimations show that this process can give a contribution comparable to the two processes
mentioned above:
$$\sigma^\mathrm{pQCD} (c c\to  c c) \sim 40 \div 200~\mathrm{nb}.$$

Thus one can conclude that predictions obtained in the LO pQCD within SPS approach underestimate the experimental value
of about $3\mathrm{\mu b}$~\cite{LHCb-PAPER-2012-003}.
Also it is worth mentioning that the experimental spectra shapes also can not be exactly reproduced.

Nevertheless some futures of the experimental spectra can be understood from such calculations using different kinematical cuts. 
For example the local minimum near 6 GeV in the experimental cross section distribution over the invariant mass $m_{cc}$ 
of two charmed particles is probably connected with the cut on the minimum transverse momenta at the LHCb data analysis~\cite{LHCb-PAPER-2012-003}: 
$$m_{cc}^\mathrm{loc.min}\approx 2p_T^\mathrm{min}.$$
Also the rapid decrease of the cross section at $m_{cc}>20$~GeV can be explained by cut on the maximum transverse momenta:  
$$m_{cc}^\mathrm{cut}\approx 2p_T^\mathrm{max}.$$

Let us now turn to the DPS contribution to the different $D$-meson pairs production.
Expression~(\ref{eqn:doubleAB}) has to be modified as experimentally observed quantities are 
inclusive production cross sections of particular types of $D$-mesons summed together with anti-mesons of
the same type. These cross sections can be written down as follows:
\begin{equation}
\sigma^{incl.}_i = \sigma_1 p^{c\vee\bar{c}}_i + \sigma_2 (2 p^{c\vee\bar{c}}_i - (p^{c\vee\bar{c}}_i)^2),
\label{eqn:}
\end{equation}
where $\sigma_1$ and $\sigma_2$ are cross sections of one and two $c\bar{c}$ pair production in a single proton-proton
collision respectively and $p^{c\vee\bar{c}}_i$ is probability that $c$ or $\bar{c}$ quark transits into detected hadron of type
$i$.

In the following we will be interested in events in which both $c$ and $\bar{c}$ quarks form two $D$-mesons
of particular type in the detector acceptance, or it is done by pairs of identical quarks --- $cc$ or $\bar{c}\bar{c}$.
In the first case cross section of the $i$ type meson pair production can be written down as
\begin{equation}
\label{eqn:DDii}
\sigma^{diff.}_{i,i} = \sigma_1 p^{c\wedge\bar{c}}_{i,i} + \sigma_2 (2 p^{c\wedge\bar{c}}_{i,i} - (p^{c\wedge\bar{c}}_{i,i})^2
+ (p^{c\vee\bar{c}}_i - p^{c\wedge\bar{c}}_{i,i})^2/2 ),
\end{equation}
and in the second --- as
\begin{equation}
\sigma^{same}_{i,i} = \sigma_2 ((p^{c\wedge\bar{c}}_{i,i})^2 
+ 2 (p^{c\wedge\bar{c}}_{i,i})(p^{c\vee\bar{c}}_i - p^{c\wedge\bar{c}}_{i,i})
+ (p^{c\vee\bar{c}}_i - p^{c\wedge\bar{c}}_{i,i})^2/2 ).
\end{equation}
Here $p^{c\wedge\bar{c}}_{i,j}$ stands for the probability for $c$ and $\bar{c}$ quarks from one pair to transit into
mesons of type $i$ and $j$ observed in the detector and probabilities for quarks from the different pairs are assumed 
independent.

For the different $i$ and $j$ types of mesons analogous quantities are written down as
\begin{eqnarray}
\sigma^{diff.}_{i,j} &=& \sigma_1 p^{c\wedge\bar{c}}_{i,j} + \sigma_2 (2 p^{c\wedge\bar{c}}_{i,j} 
- (p^{c\wedge\bar{c}}_{i,j})^2
+ 2 p^{c\wedge\bar{c}}_{i,i}p^{c\wedge\bar{c}}_{j,j} 
+ 2 p^{c\wedge\bar{c}}_{i,i} (p^{c\vee\bar{c}}_j - p^{c\wedge\bar{c}}_{i,j} - p^{c\wedge\bar{c}}_{j,j}) + 
\nonumber \\
&+& 2 p^{c\wedge\bar{c}}_{j,j} (p^{c\vee\bar{c}}_i - p^{c\wedge\bar{c}}_{i,j} - p^{c\wedge\bar{c}}_{i,i})
+ (p^{c\vee\bar{c}}_i - p^{c\wedge\bar{c}}_{i,j} - p^{c\wedge\bar{c}}_{i,i})
(p^{c\vee\bar{c}}_j - p^{c\wedge\bar{c}}_{i,j} - p^{c\wedge\bar{c}}_{j,j})),
\\
\sigma^{same}_{i,j} &=& \sigma_2 (0.5 (p^{c\wedge\bar{c}}_{i,j})^2
+ 2 p^{c\wedge\bar{c}}_{i,i}p^{c\wedge\bar{c}}_{j,j} 
+ 2 p^{c\wedge\bar{c}}_{i,i} (p^{c\vee\bar{c}}_j - p^{c\wedge\bar{c}}_{i,j} - p^{c\wedge\bar{c}}_{j,j}) +
\nonumber \\
&+& 2 p^{c\wedge\bar{c}}_{j,j} (p^{c\vee\bar{c}}_i - p^{c\wedge\bar{c}}_{i,j} - p^{c\wedge\bar{c}}_{i,i})
+ (p^{c\vee\bar{c}}_i - p^{c\wedge\bar{c}}_{i,j} - p^{c\wedge\bar{c}}_{i,i})
(p^{c\vee\bar{c}}_j - p^{c\wedge\bar{c}}_{i,j} - p^{c\wedge\bar{c}}_{j,j})).
\label{eqn:DDij}
\end{eqnarray}

To solve the equations adduced we will use known inclusive production cross sections of particular 
$D$-meson types~\cite{LHCb-CONF-2010-013}. As LHCb collaboration presents these cross sections in bins of rapidity and transverse 
momenta, they can be recalculated into the fiducial region discussed ($2<y<4$, $3\mathrm{GeV}<p_T^D<12\mathrm{GeV}$).
We will also assume that the total $c\bar{c}$ production cross section in the $7~\mathrm{GeV}$ proton-proton collisions
is known. It was obtained using the Pythia generator calibrated by known inclusive open charm production
cross sections in the LHCb acceptance and is equal to $6.1\pm0.9\mathrm{ mb}$~\cite{LHCb-CONF-2010-013}.
According to the DPS approach, cross section of two $c\bar{c}$ pairs production in a single proton-proton
scattering is given by expression (\ref{eqn:doubleAB}):
\begin{equation}
\sigma_2 = \frac{\sigma_1^2}{2 \sigma_{\rm eff.}} = 1.3\pm0.4\mbox{ mb}.
\end{equation}

However until $\sigma^{diff.}_{i,j}$ or $\sigma^{same}_{i,j}$ cross sections are measured 
there is no sufficient information to derive the $p^{c\wedge\bar{c}}_{i,j}$ probabilities. 
So we will assume that rather rigid kinematic cuts imposed result in the smallness of probability
to observe both particles produced from a $c\bar{c}$ pair in the detector. Then neglecting double
counting one can write down
\begin{equation}
p^{c\wedge\bar{c}}_{i,i} \approx (p^{c\vee\bar{c}}_i)^2, \qquad 
p^{c\wedge\bar{c}}_{i,j} \approx 2 p^{c\vee\bar{c}}_i p^{c\vee\bar{c}}_j.
\end{equation}

Under the assumptions listed equations (\ref{eqn:DDii}) --- (\ref{eqn:DDij}) can be solved.
Obtained cross sections of pair production of $D^0$, $D^+$ and $D_s^+$ mesons are 
given in Table \ref{tab:DDPS}
\begin{table}
\begin{tabular}{|c||c|c||c|c|}
\hline 
Mode                   & $\sigma^{diff.}_{th.},~\mathrm{\mu b}$ & $\sigma^{diff.}_{exp.},~\mathrm{\mu b}$ & 
$\sigma^{same}_{th.},~\mathrm{\mu b}$ & $\sigma^{same}_{exp.},~\mathrm{\mu b}$ \\ 
\hline 
$D^0 D^0$          & $7.2 \pm 1.1$   & $6.2 \pm 0.6$         & $0.53 \pm 0.2$  & $0.69 \pm 0.07$ \\ 
\hline 
$D^0 D^+$         & $6.0 \pm 0.9$    & $4.0 \pm 0.4$        & $0.4 \pm 0.1$     & $0.52 \pm 0.08$ \\ 
\hline 
$D^0 D_s^+$      & $2.3 \pm 0.4$    & $1.7 \pm 0.2$        & $0.16 \pm 0.05$   & $0.27 \pm 0.05$ \\ 
\hline 
$D^+ D^+$         & $1.2 \pm 0.2$    & $0.78 \pm 0.11$    & $0.087 \pm 0.029$ & $0.08 \pm 0.02$ \\ 
\hline 
$D^+ D_s^+$      & $0.97 \pm 0.15$  & $0.55 \pm 0.08$ & $0.066 \pm 0.022$ & $0.07 \pm 0.02$ \\ 
\hline 
$D_s^+ D_s^+$   & $0.19 \pm 0.03$  & ---                      & $0.013 \pm 0.005$  & --- \\ 
\hline 
\end{tabular} 
\caption{Cross sections of different $D$-meson pairs production compared with the LHCb results.
\label{tab:DDPS}}
\end{table}
together with the values measured by the LHCb.
We would like to stress one more time here that summation with the charge conjugate states is everywhere assumed.
Generally speaking, good agreement between the DPS predictions and the experimental results is observed. Nonetheless, 
it is mentioned in~\cite{LHCb-PAPER-2012-003} that $p_T$-spectra of $D$-mesons in pair production significantly differ 
from those in inclusive open charm production, while similar $p_T$-behaviour could be expected in the DPS model.

\section{Conclusion}

It is well known that the particle production multiplicity increases with the energy of hadronic interactions. 
Therefore phenomenon of multiple production should be observed for charmed and beauty particles as well, 
but at the higher energies due to the larger masses. 
At the LHC energy yield of charm particles (6.1~mb~\cite{LHCb-CONF-2010-013}) is comparable to the common light particle yields,
so production of two, three and so on pairs should be expected as well as single $c\bar{c}$ pair production.
Recently the first data on the four $c$-quark production in the proton-proton interactions have been obtained by
the LHCb Collaboration~\cite{LHCb-PAPER-2012-003}. 

From the theoretical point of view processes in single gluon-gluon interactions (such as $gg \to c\bar{c}c\bar{c}$) 
are the natural source of multiple charm production. The calculations within LO of pQCD in SPS approach had been done 
earlier for the process of $J/\psi$ pair production~\cite{Humpert:1983yj,Ko:2010xy,Berezhnoy:2011xy},
$J/\psi+c\bar{c}$ associated production~\cite{Barger:1991vn, Berezhnoy:1998aa, Baranov:2006dh, 
Baranov:2006rz, Artoisenet:2007xi, He:2009zzb}, and for the four $c$-quarks production.

The main conclusion to be drawn from these theoretical studies and from the recent LHCb results is that SPS model 
used together with the LO pQCD can not describe all the data on multiple charm production. 
The presented analysis shows that only data on $J/\psi$ pair production is in satisfactory agreement with SPS LO pQCD predictions. 
The predictions obtained for the $J/\psi + D$ associated production, as well as for the four $D$-meson production 
underestimate the experimental data in several times. As alternative model we consider the simplest model of double parton scattering 
(DPS). In the frame work of this approach it is assumed that two $c\bar{c}$ pairs are produced independently in
two different partonic collisions. DPS predictions on the cross section values fairly agree with the experimental data. 
As it was shown in~\cite{Luszczak:2011zp}, cross section of pair charm production becomes equal to the ordinary
$c\bar{c}$ cross section at the energy of about 20~TeV.

It is interesting to note, that for the double $J/\psi$ production predictions of SPS and DPS models are fairly close, for the $J/\psi + D$
associated production the DPS prediction exceeds the SPS one in several times and for the four $D$-meson production excess is even higher.
At first glance it seems amazing as an attempt to explain advantage of the DPS model by combinatorial factor only
does not lead to distinction in the channels discussed. Infinitesimality connected with the $\alpha_S$ constant is same for both
SPS and DPS: in SPS the factor is $\alpha_S^4$ and in DPS --- $\alpha_S^2 \times \alpha_S^2 = \alpha_S^4$. 
From our point of view the reasonable explanation lies in the different phase volumes for the SPS and DPS production: 
in SPS final state contains three particles for the $J/\psi + c \bar{c}$ production and four for the $c \bar{c} c \bar{c}$
production, so differential cross sections of these processes peak at the larger $\sqrt{\hat{s}}$ values
at the expense of phase volume factors. By-turn this leads to the smaller gluon luminosity as compared to the $2 \to 2$ processes
which take place in the DPS model.

Authors would like to thank Vanya Belyaev for fruitful discussions. The work was financially supported by 
Russian Foundation for Basic Research (grant \#10-02-00061a), grant of the president of Russian Federation 
(\#MK-3513.2012.2) and by the non-commercial foundation ``Dynasty''.

\bibliography{multicharm}

\end{document}